\documentclass[a4paper,twoside]{article}

    \usepackage{amsmath,amssymb}
    \usepackage{bm}
    \numberwithin{equation}{section}

\begin{document}

    \setlength{\baselineskip}{6mm}

\begin{center}

\vspace*{10mm}

{\LARGE \textbf{An Attempt to Remove Quadratic}
\vspace*{2mm}

\textbf{Divergences in the Standard Theory}}

\vspace*{10mm}
{\large Noboru NAKANISHI}
 \footnote{Professor Emeritus of Kyoto University. E-mail: nbr-nak@trio.plala.or.jp}\\

\textit{12-20 Asahigaoka-cho, Hirakata 573-0026, Japan}

\end{center}

\vspace*{5mm}

    \setlength{\baselineskip}{5mm}

{\small The quadratic divergences caused by Yukawa interactions in the 
standard theory of elementary particle physics is shown to be removed by
introducing finite-mass complex-ghost regulator fields. In this modification
of the standard theory, its
manifest covariance, renormalizability, gauge invariance and unitarity are
retained, and no new observable particles are introduced.}

    \setlength{\baselineskip}{6mm}

\section{Introduction}

The standard theory of elementary particle physics (electroweak theory plus quantum
chromodynamics) is a very successful theory.
It is formulated as a \textit{local} quantum field theory in 4-dimensional spacetime 
$^{1), 2)}$ and its predictions have no clear contradictions with high-energy 
experimental results,
provided that right-handed neutrinos are taken into account.  
This theory has the following fundamental properties:

1. Its Lagrangian density is manifestly covariant.

2. It is renormalizable.

3. It is invariant under $SU(3)_c\times SU(2)_L \times U(1)_Y$ gauge transformations.

4. Its physical S-matrix is unitary, though indefinite metric is used for BRS
quantization 

\hspace*{4mm} of gauge fields.
   
The $SU(2)_L \times U(1)_Y$ gauge invariance is spontaneously 
broken up to $U(1)_{em}$ by Higgs mechanism. Higgs field has four real
field-degrees of freedom; three of them are Nambu-Goldstone bosons, which are
unphysical owing to the Kugo-Ojima subsidiary condition,
and the remaining one is a massive scalar boson called Higgs boson.
In order to give nonzero masses to leptons and quarks, it is assumed that
there is a Yukawa interaction between Higgs field and each of them;
the mass is essentially a product of the Yukawa coupling constant and the
vacuum expecation value of Higgs field. This Yukawa interaction causes 
quadratic divergences in the self-energy part of Higgs boson.  

Because the standard theory is renormalizable, the renormalized S-matrix is finite
to all orders of perturbation theory. But if one wishes to discuss radiative
masses, one must introduce a cutoff parameter $\varLambda$. If $\varLambda$
is taken to be of order of Planck mass, quadratically divergent quantities 
give uncontrollably large contributions. This trouble is known as the 
hierarchy problem. It may be resolved by the supersymmetric theory (SUSY),
but since no superparticles are not yet observed at all, it is quite unlikely
that SUSY is realized in Nature. 

The purpose of the present paper is to propose
a possible modification of the standard theory for removing the quadratic 
divergences caused by the Yukawa interactions,
in such a way that the above-mentioned fundamental properies are retained
and that no new observable particles are necessary to be introduced.  
As is well known, ultraviolet divergences can be removed by the regulator method
(though it is quite nontrivial to remove all divergences so as to be consistent with 
non-abelian gauge invariance), but the introduction of indefinite
metric usually violates the unitarity of the physical S-matrix. As was 
emphasized previously,$^{3), 4)}$ however, complex-ghost 
quantum field theory does not violate
the unitarity, because complex ghosts cannot appear \footnote{Precisely speaking,
the probability with which 
a pair of a complex ghost and its complex-conjugate ghost appear is of measure 
\textit{zero}.} in the final state if they are
absent in the initial state, owing to energy conservation law.
Although Lorentz invariance of the S-matrix is known to be violated in this theory,
the amount of the violation can be made small by appropriately choosing a 
certain parameter that characterizes the imaginary part of the complex-ghost mass.
Accordinly, it is quite an attractive possibility to use complex ghosts as regulators.
In the present paper, we show that the complex-ghost regulater method can remove
the quadratic divergences caused by Yukawa interactions without violating the
above-mentioned fundamental properties of the standard theory.

The present paper is organized as follows.
In \S 2, the standard theory is very briefly reviewed.  In \S 3, we present the
theory of complex-ghost regulators. In \S 4, we show how the quadratic
divergences caused by Yukawa interactions can be removed by the 
complex-ghost regulators.

\section{Standard theory}

The Lagrangian density of the standard theory consists of gauge-field part, lepton
part, quark part and Higgs part. In the present paper, as we discuss Yukawa
interactions,  lepton part and quark part only are explicitly considered. 

The lepton-part Lagrangian density is written as a sum over three generations 
$\alpha = e, \mu, \tau$. In each generation, the left-handed charged 
lepton $\alpha_L$ and the left-handed neutrino $\nu_{\alpha L}$
constitute an $SU(2)_L$ doublet $l_\alpha$ with a hypercharge $Y=-1/2$, 
while the right-handed charged lepton $\alpha_R$ constitutes an $SU(2)_L$
singlet with $Y=-1$. Furthermore, recent neutrino oscillation experiments$^{5)}$
require the introduction of the right-handed neutrino $\nu'_{\alpha R}$,
which is an $SU(2)_L$ singlet with $Y=0$, and  of the unitary neutrino-mixing
matrix $U = (U_{\alpha \beta})$. We denote  $\sum_\beta
U_{\alpha \beta} \nu_\beta$
by $\tilde{\nu_\beta}$ and $l_\alpha$ with $\tilde{\nu_\alpha}$
in place of $\nu_\alpha$ by $\tilde{l}_\alpha$; we set $\hat{l}_\alpha
= (U^{-1}\tilde{l})_\alpha$.
The Higgs field $\varPhi$ is an $SU(2)_L$ doublet with $Y=1/2$, and 
we set $\tilde{\varPhi} = i\tau_2 \varPhi^*$.

The lepton-part Lagrangian density is given by
\begin{equation} 
 \begin{split}
\mathcal{L}_{\mathrm{lepton}} = & \sum_{\alpha=e, \mu, \tau}
\bigl[\mspace{1mu}
\overline{\tilde{l}}_\alpha i\gamma^\mu \mathcal{D}^{L+Y}_\mu \tilde{l}_\alpha +
\overline{r}_\alpha i\gamma^\mu \mathcal{D}^Y_\mu r_\alpha +
\overline{r}'_\alpha i\gamma^\mu \partial_\mu r'_\alpha \\
+ & (-f_\alpha)\{(\overline{\hat{l}}_\alpha \varPhi) r_\alpha +
\overline{r}_\alpha (\varPhi^\dagger \hat{l}_\alpha)\}
+ (-f'_\alpha)\{(\overline{\tilde{l}}_\alpha \tilde{\varPhi}) r'_\alpha +
\overline{r}'_\alpha (\tilde{\varPhi}^\dagger \tilde{l}_\alpha)\} \bigr],
 \end{split}
\end{equation}
where $\mathcal{D}^{L+Y}_\mu$ and $\mathcal{D}^Y_\mu$ denote covariant
differentiations with respect to $SU(2)_L \times U(1)_Y$ and $U(1)_Y$,
respectively.

The quark-part Lagrangian density has essentially the same form as the
lepton-part one does, except for the fact that the former has the color degrees
of freedom, which is inessential to the present work. The ``upper" quarks
\{$u$, $c$, $t$\} and the ``lower" quarks \{$d$, $s$, $b$\} correspond
to the charged antileptons \{$\overline{e}$, $\overline{\mu}$, $\overline{\tau}$\}
and to the antineutrinos \{$\overline{\nu}_e$, $\overline{\nu}_\mu$, 
$\overline{\nu}_\tau$\}, respectively.
In this correspondence, the values of the hypercharge should be changed as
$Y_\mathrm{quark} = Y_\mathrm{antilepton} -1/3$.

The lower component of the Higgs field $\varPhi$ acquires a nonvanishing
vacuum expectation value $v/\sqrt{2}$ \hspace{1mm}($v>0$); 
hence the vacuum expectation
value of the upper component of $\tilde{\varPhi}$ is also $v/\sqrt{2}$. If
$\varPhi$ is reexpressed in terms of $v$ and four hermitian fields (Higgs boson
and 3-component Nambu-Goldstone boson), the quadratic part of 
$\mathcal{L}_{\mathrm{lepton}}$ becomes the free Dirac Lagrangian density
for all leptons, modified by neutrino mixing.
 The mass $m_\alpha$ of a charged lepton $\alpha$ is given by $f_\alpha
v/\sqrt{2}$. As for neutrinos, we encounter neutrino-mixing mass matrix,
$\mathcal{M}$. Likewise for $\mathcal{L}_{\mathrm{quark}}$.

\section{Complex-ghost regulators}

We introduce pairs of Weyl-spinor fields $L_j$ ($SU(2)_L$ doublet, $Y=-1/2$), 
$R_j$ ($SU(2)_L$ singlet, $Y=-1$) and $R'_j$ ($SU(2)_L$ singlet, $Y=0$) ($j=1,2$); 
the $j=1$ fields have positive norm, while the $j=2$ ones have negative 
norm. Imitating (2.1), we introduce the Lagrangian
density
\begin{equation} 
 \begin{split}
\mathcal{L}_{\mathrm{regulator}} = & \sum_{j=1,2} (-1)^{j-1}
\bigl[
\overline{L}_j i\gamma^\mu \mathcal{D}^{L+Y}_\mu L_j +
\overline{R}_j i\gamma^\mu \mathcal{D}^Y_\mu R_j +
\overline{R}'_j i\gamma^\mu \partial_\mu R'_j \bigr] \\
+ & \sum_{j,k=1}^2 \bigl[
(-f_{jk})\{(\overline{L}_j \varPhi) R_k +
\overline{R}_j (\varPhi^\dagger L_k)\}
+ (-f'_{jk})\{(\overline{L}_j \tilde{\varPhi}) R'_k +
\overline{R}'_j (\tilde{\varPhi}^\dagger L_k)\} \bigr],
 \end{split}
\end{equation}
where $f_{jk} = f_{kj}^*$ and $f'_{jk} = {f'_{kj}}^*$.

$\varPhi$ is reexpressed in terms of $v$ and four real fields.
We set
\begin{equation}
 \begin{split}
f_{11} v/\sqrt{2} = m_1, \mspace{10mu}
f_{22} v/\sqrt{2} = -m_2, \mspace{10mu}
f_{12} v/\sqrt{2} = \gamma/2; \\
f'_{11} v/\sqrt{2} = m'_1, \mspace{10mu}
f'_{22} v/\sqrt{2} = -m'_2, \mspace{10mu}
f'_{12} v/\sqrt{2} = \gamma'/2.
 \end{split}
\end{equation}
Then the quadratic part of (3.1) becomes
\begin{equation}
 \begin{split}
\mathcal{L}_{\mathrm{regulator}}^0
= & \sum_{j=1}^2 (-1)^{j-1} (\overline{\varPsi}_j i\gamma^\mu
\partial_\mu \varPsi_j -m_j \overline{\varPsi}_j \varPsi_j)
-\frac{\gamma}{2} \overline{\varPsi}_1 \varPsi_2
-\frac{\gamma^*}{2} \overline{\varPsi}_2 \varPsi_1 \\
+ & \sum_{j=1}^2 (-1)^{j-1} (\overline{\varPsi}'_j i\gamma^\mu
\partial_\mu \varPsi'_j -m'_j \overline{\varPsi}'_j \varPsi'_j)
-\frac{\gamma'}{2} \overline{\varPsi}'_1 \varPsi'_2
-\frac{{\gamma'}^*}{2} \overline{\varPsi}'_2 \varPsi'_1,
 \end{split}
\end{equation}
where $\varPsi_j$ is composed of the upper component of $L_j$ and $R_j$
and $\varPsi'_j$ is composed of the lower component of $L_j$ and $R'_j$.

The field equations derived from (3.3) are
\begin{equation}
 \begin{split}
(i\gamma^\mu \partial_\mu -m_1)\varPsi_1 - \frac{\gamma}{2} \varPsi_2  & =0, \\
(i\gamma^\mu \partial_\mu -m_2)\varPsi_2 + \frac{\gamma^*}{2} \varPsi_1  & =0.
 \end{split}
\end{equation}

Canonical quantization is performed with \textit{``wrong statistics"},
that is, we set up \textit{commutation relations} but not anticommutation
relations:
\begin{equation}
 \begin{split}
[\varPsi_j (x), \mspace{2mu} \overline{\varPsi}_k (y)]_{x_0 = y_0}  & =
(-1)^{j-1} \gamma^0 \delta (\bm{x}-\bm{y}) \mspace{20mu} (j=k) \\ 
& =0 
\mspace{20mu} (j \neq k).
 \end{split}
\end{equation}

Then we set up a Cauchy problem for the 4-dimensional commutators
\begin{equation}
[\varPsi_j (x), \mspace{2mu} \overline{\varPsi}_k (y)] \equiv iS_{jk} (x-y).
\end{equation}
As was done previously,$^{6), 7)}$ the solution to the Cauchy problem is easily found
by diagonalizing the mass matrix
\begin{equation}
\bm{M} = \left(
\begin{array}{cc}
m_1 & \gamma/2 \\
-\gamma^*/2 & m_2
\end{array}
\right).
\end{equation}
The eigenvalues become complex if
\begin{equation}
|\gamma| > |m_1 - m_2|.
\end{equation} 
Then we obtain the complex masses $M$ and $M^*$, where
\begin{equation}
M \equiv \frac{m_1 + m_2}{2} + \frac{i}{2}\sqrt{\gamma \gamma^* -
(m_1 - m_2)^2}.
\end{equation}
Hence,
\begin{equation}
M^2 + M^{*2} = m_1^2 + m_2^2 - \frac{\gamma\gamma^*}{2}
=\mathrm{\bm{tr}} \bm{M^2}.
\end{equation}

We can then construct the Wightman functions $S_{jk}^{(+)}(x-y)$
and the Feynman propagator $S_{Fjk}(x-y)$ quite analogously to the
case of the scalar complex ghost. But we do not need their explicit
expressions for the present purpose.

The same procedure as above is carried out for primed quantities.
That is, corresponding to (3.4) - (3.10),
the equations in which $\varPsi_j$, $m_j$, $\gamma$, $S_{jk}$, $\bm{M}$
and $M$ are replaced by the respective primed ones hold.

 Furthermore, for the quark part, everything goes quite analogously to the above
consideration on the lepton part. Of course,
in the quark case, the complex-ghost regulators have the color degrees of freedom.

\section{Removal of quadratic divergences}

As is well known, the appearance of quadratic divergences in the standard theory
is only in the proper self-energy part of Higgs boson. In the present
paper, we discuss the quadratic divergences caused by the Yukawa interaction.
That is, we consider only the self-energy Feynman diagrams in which the end vertices
of both external Higgs-boson lines correspond to Yukawa interactions.

The internal-line part of such a Feynman diagram consist of a lepton or quark loop
together with radiative corrections. Because of the Fermi statistics of leptons 
and quarks, this loop
gives an overall factor $-1$ to the Feynman integral.
On the other hand, \textit{the Feynman diagram that has a complex-ghost regulator loop
acquires no such a factor because of the Bose statistics of the regulator
fields}. Furthermore, it is important to note that as far as 
quadratic divergences are concerned,
\textit{the mass term of any Feynman propagator is irrelevant}; that is, we may 
forget about the mass except for the fact that each Yukawa coupling constant
is proportional to a mass (or mass matrix element).

We first consider the second-order self-energy parts.
The contribution from the charged lepton-loop diagrams is proportional to
$-\sum_\alpha m_\alpha^2$. The contribution from the corresponding
regulator-loop diagrams is proportional to 
\begin{equation}
f_{11}^2 + 2(-1)f_{12} f_{21}+ (-1)^2 f_{22}^2 = \left(\frac{v}{\sqrt{2}}\right)^{-2}
\left( m_1^2 + m_2^2 -\frac{\gamma\gamma^*}{2} \right).
\end{equation}
Here the factor $-1$ is due to the indefinite metric appearing in (3.3).
Hence the quadratic divergence caused by three charged leptons is cancelled if 
\begin{equation}
\sum_\alpha m_\alpha^2 = M^2 + M^{*2}
\end{equation}
Likewise, the quadratic divergence caused by three neutrinos is cancelled if
\begin{equation}
\mathrm{tr} \mathcal{M}^2 = {M'}^2 + {M'}^{*2},
\end{equation}
where $\mathcal{M}$ denotes the neutrino-mixing mass matrix.

Unfortunately, for higher-order self-energy diagrams, such
a simple cancellation condition as above is no longer valid.
In order to realize the removal of quadratic divergences to all orders,
we must introduce complex-ghost regulator fields for
\textit{each} generation.
That is, we must replace (3.1) by
\begin{equation} 
 \begin{split}
\mathcal{L}_{\mathrm{regulator}} = \sum_{\alpha=e,\mu,\tau} 
\Bigl( & \sum_{j=1,2} (-1)^{j-1} \bigl[
\overline{\tilde{L}}_{\alpha j} i\gamma^\mu \mathcal{D}^{L+Y}_\mu 
\tilde{L}_{\alpha j} +
\overline{R}_{\alpha j} i\gamma^\mu \mathcal{D}^Y_\mu R_{\alpha j} +
\overline{R}'_{\alpha j} i\gamma^\mu \partial_\mu R'_{\alpha j} \bigr] \\
+ & \sum_{j,k=1}^2 \bigl[
(-f_{\alpha jk})\{(\overline{\hat{L}}_{\alpha j} \varPhi) R_{\alpha k} +
\overline{R}_{\alpha j} (\varPhi^\dagger \hat{L}_{\alpha k})\} \\
 & \mspace{20mu} +(-f'_{\alpha jk})\{(\overline{\tilde{L}}_{\alpha j} 
\tilde{\varPhi}) R'_{\alpha k} +
\overline{R}'_{\alpha j} (\tilde{\varPhi}^\dagger \tilde{L}_{\alpha k})\} 
\bigr] \Bigr),
 \end{split}
\end{equation}
where notation will be obvious from the above consideration.
 
Then, for example, for a charged lepton $\alpha$,
(4.2) should be replaced by 
\begin{equation}
m_\alpha^2 = M_\alpha^2 + M_\alpha^{*2} \mspace{20mu} 
\mathrm{for} \mspace{5mu} \mathrm{each} \mspace{5mu}  \alpha, 
\end{equation}
where $M_\alpha$ denotes the complex mass acquired by $\varPsi_{\alpha j}$.
As for the neutrino-loop self-energy part, a similar relation should be set up
for each eigenvalue of the mass matrix.

For the quark part, everything is analogous to the above. 

Thus, the quadratic divergences caused by Yukawa interactions can be removed
by the introduction of the complex-ghost regulator fields without violating
the unitarity of the physical S-matrix, though Lorentz invariance is spontaneously
violated slightly.

\vspace*{5mm}

{\small
\begin{center}
\textbf{References}\\
\end{center}

1) \; T. Kugo, \textit{Quantum Theory of Gauge Fields, II} (Baifukan, 1989) (in Japanese).

2) \; K. Aoiki, Z. Hioki, R. Kawabe, MDKonuma and T. Muta, Prog.~Theor.~Phys.~Suppl. No.73 

\hspace*{8mm} (1982), 1.

3) \; N. Nakanishi, Prog.~Theor.~Phys.~Suppl. No.51 (1973), 1. Further 
references are contained 

\hspace*{8mm} therein.

4) \; N. Nakanishi, Phys.~Rev.~\textbf{D5} (1972), 1968.

5) \; S. Eidelman \textit{et al}, Phys. Letters ~\textbf{B592} (2004), 1, Chap.15.

6) \; N. Nakanishi, Prog.~Theor.~Phys.~\textbf{116} (2006), 873.

7) \; N. Nakanishi, Prog.~Theor.~Phys.~\textbf{118} (2007), 913.

\end{document}